\documentclass[10pt]{article}

\usepackage{arxiv}

\usepackage[utf8]{inputenc} % allow utf-8 input
\usepackage[T1]{fontenc}    % use 8-bit T1 fonts
\usepackage[hidelinks]{hyperref}       % hyperlinks
\usepackage{url}            % simple URL typesetting
\usepackage{booktabs}       % professional-quality tables
\usepackage{amsfonts}       % blackboard math symbols
\usepackage{nicefrac}       % compact symbols for 1/2, etc.
\usepackage{microtype}      % microtypography
\usepackage{xcolor}
\usepackage{graphicx}
\usepackage{subcaption}
\usepackage{multicol}
\usepackage[ruled,vlined]{algorithm2e}
\usepackage{cite}
\usepackage{titling}
\usepackage{caption}

\newenvironment{Figure}
  {\par\medskip\noindent\minipage{\linewidth}}
  {\endminipage\par\medskip}

\title{MiNAA: Microbiome Network Alignment Algorithm}

\author{
Reed Nelson \\
Department of Computer Science \\
University of Wisconsin-Madison
\and 
Rosa Aghdam\\
Wisconsin Institute for Discovery \\
University of Wisconsin-Madison
\and 
Claudia Sol\'is-Lemus\thanks{Corresponding author: solislemus@wisc.edu}\\
Wisconsin Institute for Discovery \\
Department of Plant Pathology \\
University of Wisconsin-Madison
}
\date{}

\begin{document}
\maketitle

%\begin{abstract*}
%\textbf{Summary:} 
%\textbf{Availability and Implementation:} MiNAA is a freely available software implemented in C++ with all major operating systems supported. The code is publicly available and open-source (\url{https://github.com/solislemuslab/minaa}).
%\textbf{Contact:} solislemus@wisc.edu
%\textbf{Supplementary information:} Supplementary data are available at \textit{Bioinformatics} online.
%\end{abstract*}

% keywords can be removed
%\keywords{xxx \and xxxg \and xxx \and Uxxxs \and xxx}

%\begin{multicols}{2}
\section{Summary}

A microbial network is a mathematical representation of a microbial community where nodes represent microbes and edges represent interactions. It is well-recognized that microbes are among the main drivers of biological phenotypes in soil, plants, and animals alike, and thus, their study has vast implications for soil, plant and human health. In particular, recognizing the microbial, environmental, and agricultural factors that drive plant and soil phenotypes is crucial to comprehend processes connected to plant and soil health, to identify global practices of sustainable agriculture, as well as to predict plant and soil responses to environmental perturbations such as climate change.

The adaptability of microbes to thrive in every environment poses challenges for scientists who try to understand their communities. Indeed, two microbial communities with the exact same players can interact differently depending on the environmental conditions. It is thus desirable to identify commonalities and differences on two microbial networks, hence the need for computational tools that can match (or \textit{align}) them.

\section{Statement of need}
Our Microbiome Network Alignment Algorithm (\texttt{MiNAA}) aligns two microbial networks using a combination of the GRAph ALigner (\texttt{GRAAL}) algorithm \cite{kuchaiev2010topological} and the Hungarian algorithm \cite{kuhn1955hungarian, Pilgrim1995}. Network alignment algorithms find pairs of nodes (one node from the first network and the other node from the second network) that have the highest \textit{similarity}. Traditionally, similarity has been defined as topological similarity such that the neighborhoods around the two nodes are similar. Recent implementations of network alignment methods such as \texttt{NETAL} \cite{neyshabur2013netal} and \texttt{L-GRAAL} \cite{malod2015graal} also include measures of biological similarity, yet these methods are restricted to one specific type of biological similarity (e.g. sequence similarity in \texttt{L-GRAAL}). Our work extends existing network alignment implementations by allowing \textit{any} type of biological similarity to be input by the user. This flexibility allows the user to choose whatever measure of biological similarity is suitable for the study at hand.
In addition, unlike most existing network alignment methods that are tailored for protein or gene interation networks \cite{chen2020gppial,ma2020review}, our work is the first one suited for microbiome networks.

\section{Description of the \texttt{MiNAA} algorithm}

\noindent \textbf{Input.}
Two networks, represented by adjacency matrices, are the main inputs for our algorithm. Let $G$ and $H$ be such input networks such that $|G| \leq |H|$, where $|G|$ indicates the size of $G$. Optionally, the user may add biological similarity (as a $|G| \times |H|$ matrix) to be weighed into the alignment. This matrix could include gene similarity, phylogenetic similarity, functional similarity, among others. Our algorithm also includes additional options that allow the user to specify how much weight should be placed on biological versus topological information. Specifics on all the input arguments can be found on GitHub \url{https://github.com/solislemuslab/minaa}.

\noindent \textbf{Algorithms.}
For each input network ($G, H$), we calculate the \textit{graphlet degree vector}, also denoted the \textit{node signature} of each node. This topological descriptor characterizes a node based on its local neighborhood within a 5-node radius \cite{prvzulj2007biological}. 

Currently, we use the same algorithm for calculating node signatures as in \texttt{GraphCrunch2} \cite{kuchaiev2011graphcrunch} which is a $O(ed^3)$ algorithm where $e$ is the number of edges, and $d$ is the maximal node degree. Future work will see this implementation replaced by ORCA \cite{hovcevar2016computation}, a functional equivalent with runtime $O(ed^2)$ for sparse graphs such as microbial networks.

Using the node signatures, we calculate the topological difference between each node $g \in G$ and $h \in H$, according to the formula described in \cite{milenkovic2008uncovering}. In keeping with the flexibility of the original formula, we include a parameter $\alpha$, such that the topological distance $T_{i,j} = \alpha \cdot S_{i,j} + (1 - \alpha) \cdot N_{i,j}$, where $S_{i,j}$ represents the difference in node signatures between nodes $i$ and $j$, and $N_{i,j}$ represents the difference in degree between the nodes $i$ and $j$. It is recommended to use the default $\alpha = 1$. We store the resulting values in the topological cost matrix $T$ whose computation has complexity $O(|G||H|)$.

At this time, if the user provided a biological cost matrix $B$, the two matrices are combined into an overall cost matrix $O$, such that $O_{i,j} = \beta \cdot T_{i,j} + (1 - \beta) \cdot B_{i,j}$. The user can specify the value of the parameter $\beta$, and by default, we use $\beta=1$ which represents the case in which only topological information is considered.

The overall cost matrix, $O$, is the input to the Hungarian algorithm \cite{kuhn1955hungarian, Pilgrim1995}, an $O(|G|^3)$ solution to the \textit{assignment problem}. It is important to note that the Hungarian algorithm will align every node $G$ to some node in $H$. Optimally, we should only align nodes which we have sufficient confidence ought to be aligned, but this \emph{partial assignment problem} is computationally harder (likely, NP hard), and beyond the scope of this work.
We show a graphical abstract of our algorithm in Figure \ref{graph-abs}.

\begin{Figure}
   \centering
    \includegraphics[width=15cm]{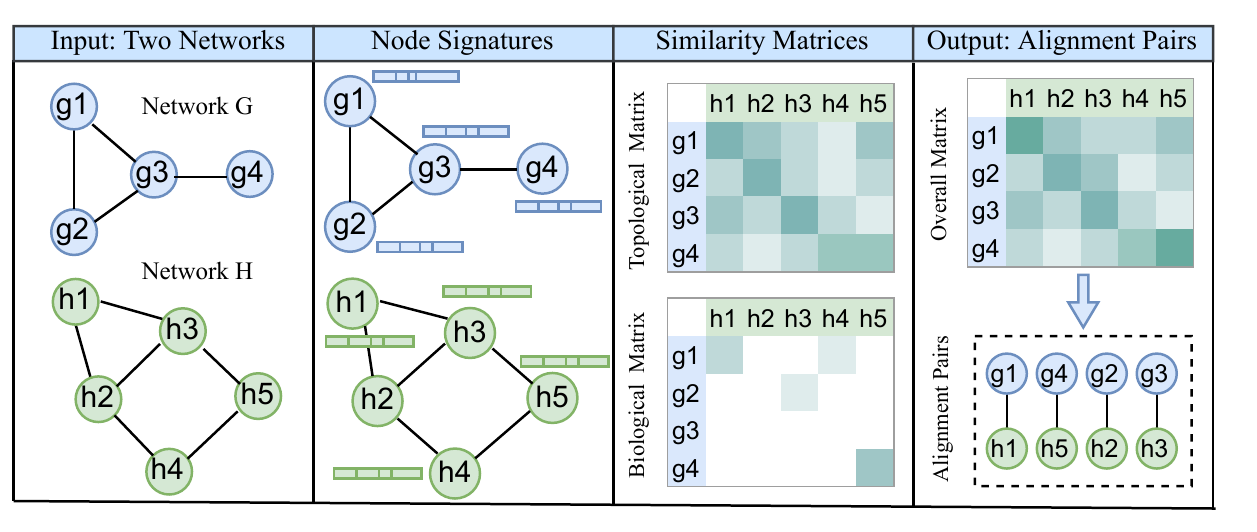} 
    \captionof{figure}{Graphical abstract of \texttt{MiNAA}. The algorithm takes two networks as input ($G,H$), and for each node in the networks, it computes a node signature vector based on the topological information of the node's neighborhood. With the node signatures, a topological cost matrix is computed which is then combined with a biological cost matrix input by the user. This method then applies the Hungarian algorithm to the overall cost matrix, resulting in a globally optimal alignment. Note that we show \textit{similarity} matrices rather than \textit{cost} matrices in this figure. While the algorithm works with cost matrices, similarity matrices are more intuitive for the graphical representation, and it is easy to convert one for the other with a transformation.}
    \label{graph-abs}
\end{Figure}

\noindent \textbf{Output.}
The algorithm's main output is the complete alignment of $G$ and $H$. That is, we return a list of pairs of nodes, one node from network $G$ and one node from network $H$ that have been identified by the algorithm as having the lowest alignment cost. We also return an alignment score matrix with dimensions $|G| \times |H|$ with the alignment score of each pair of nodes (the highest value, the more evidence for alignment). Along the way, $G$ and $H$'s node signature sets are saved as files, as are the intermediate matrices $T$, $B$, and $O$.

\section{Simulations}

We simulate networks with the R package \texttt{SPIEC-EASI} (SParse InversE
Covariance Estimation for Ecological Association Inference) \cite{kurtz2015sparse}.
We focus on the ``Cluster" network topology since Cluster networks are conducive to speculative ecological scenarios. Indeed, for microbial communities that inhabit many discontinuous niches (clusters) and have minimal interactions between niches, cluster networks may serve as archetypal models  \cite{peschel2021netcomi,yang2019meta}.

A network is simulated with certain number of nodes, and then a proportion of its edges are flipped to produce a second network. Because the second network is just a perturbation of the first network, we expect our alignment method to align the same nodes correctly.
For example, for the original simulated network (Network $G$) with 10 nodes and with 5\% edge change rate, five out of every 100 edges in the adjacency matrix that are 1 are replaced by 0, and similarly, five out of every 100 edges that are 0 are replaced with 1. The modified network from Network $G$ is named Network $H$. Then our algorithm is applied on the original and modified networks to detect the alignment pairs. This results in a alignment matrix with its elements represented as $a_{ij}$. Each $a_{ij}$ shows the similarity value of the $i$'s node in Network $G$ with $j$'s node in Network $H$. We expect the alignment matrix to be close to the identity matrix.

We iterate over different numbers of nodes (10, 30, 50, 100, 250 and 500), and different proportions of edges changed (0.05 (5\%), 0.1, and 0.9). 
To achieve reliable results, each scenario is repeated 30 times and the average of the alignment matrices is computed. We expect Network $G$ to be properly aligned with a perturbation of itself with edge change rate of 0.05 or 0.1, but not with an edge change rate of 0.9.

\begin{figure}[h]
\centering
\includegraphics[width=18cm]{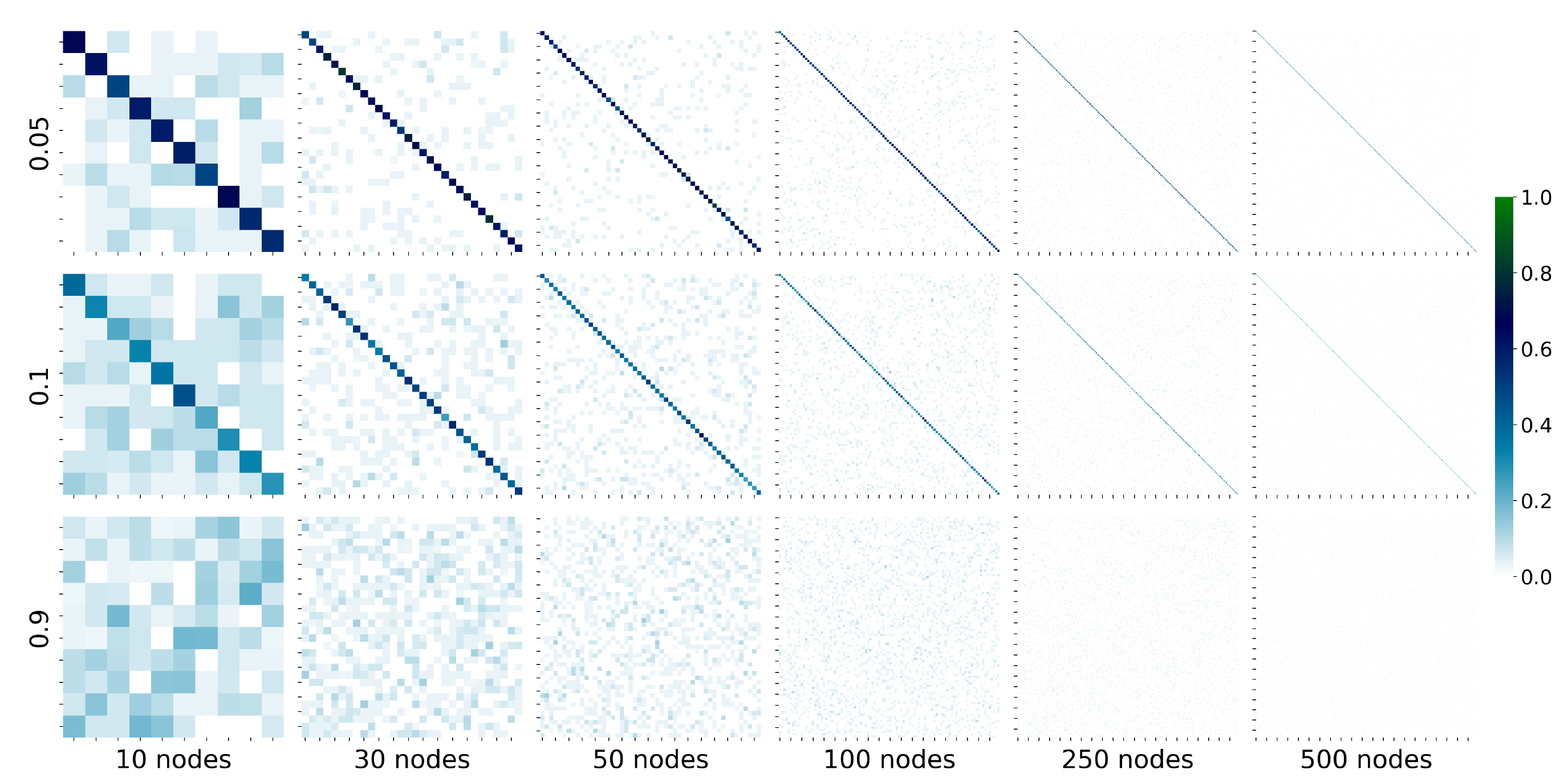}
\captionof{figure}{Averaged alignment matrices for all simulated networks, rows representing the rate of altered edge change and columns representing the number of nodes. As expected, for 0.05 or 0.1 edge change rate, the alignment matrices are close to identity matrices illustrating our method's ability to align the same nodes on perturbed networks.}\label{heatmap}
\end{figure}

Figure \ref{heatmap} displays the results of the averaged similarity matrices for all simulated networks, with rows representing the edge change rate and columns representing the number of nodes. For instance, for 10 nodes with 0.05 edge changes, the heatmap displays the mean of 30 alignment matrices. Given that we observe diagonal matrices for a small rate of edge change (0.05 and 0.1), we can conclude that our algorithm correctly aligns nodes under small perturbations of the networks. However, the results for the same networs with the edge change rate of 0.9 is far from diagonal which is also what we expected.

We also present results on running time (Table \ref{tabletime}).
We average the runtime over 90 alignments (30 replicates with 3 levels of edge change rate each) for each network size (as in the simulated data described above). Benchmarking was done in a single thread on a 72 core/3.10GHz Intel CPU, with 1TB available RAM.

\begin{table}[h]
    \centering
    \begin{tabular}{|c|c|c|c|c|c|c|}
     \hline
     Number of nodes & 10 & 30 & 50 & 100 & 250 & 500 \\
     \hline
     Average Runtime (ms) & 25.100 & 40.167 & 66.767 & 139.133 & 972.433 & 7603.233 \\
     \hline
     \end{tabular}
    \caption{Running time (in ms) of the average over 90 alignments (30 replicates with 3 levels of edge change rate each). Even the alignment of the networks with 500 nodes takes less than 8 seconds.}
    \label{tabletime}
\end{table}

%\noindent \textbf{Open-source code.} All of our code is open source and available in the following GitHub repository \url{https://github.com/solislemuslab/minaa}.

\section{Acknowledgements} 
This work was supported by the Department of Energy [DE-SC0021016 to CSL]. The authors thank Arnaud Becheler for meaningful discussions about ORCA.

%\footnotesize{
\bibliographystyle{unsrt}  
\bibliography{references}  %%% Remove comment to use the external .bib file (using bibtex).

%\end{multicols}
\end{document}